\documentclass[submission, Proceedings]{SciPost}

\hypersetup{
    colorlinks,
    linkcolor={red!50!black},
    citecolor={blue!50!black},
    urlcolor={blue!80!black}
}

\usepackage[bitstream-charter]{mathdesign}
\usepackage{xspace}
\usepackage{caption}
\usepackage{subcaption}

\DeclareSymbolFont{usualmathcal}{OMS}{cmsy}{m}{n}
\DeclareSymbolFontAlphabet{\mathcal}{usualmathcal}

\newcommand{\be}{\begin{equation}}
\newcommand{\ee}{\end{equation}}
\newcommand{\pom}{{I\!\!P}}
\newcommand{\regg}{{I\!\!R}}\newcommand\Eq[1]{(\ref{#1})}
\newcommand\Fig[1]{Fig.~\ref{#1}}
\newcommand\Sec[1]{Sec.~\ref{#1}}

\DeclareRobustCommand\GeV{\ensuremath{\mathrm{GeV}}\xspace}
\DeclareRobustCommand\sred{\ensuremath{\sigma_\mathrm{r}}\xspace}
\newcommand\DD{\mathrm{D}}
\DeclareRobustCommand\FL{\ensuremath{F_\mathrm{L}}\xspace}
\DeclareRobustCommand\xL{\ensuremath{x_\mathrm{L}}\xspace}
\DeclareRobustCommand\estat{\ensuremath{\delta_\mathrm{stat}}\xspace}
\DeclareRobustCommand\esys{\ensuremath{\delta_\mathrm{sys}}\xspace}
\DeclareRobustCommand\pT{\ensuremath{p_\perp}\xspace}
\DeclareRobustCommand\Ep{\ensuremath{E_p}\xspace}
\DeclareRobustCommand\Ee{\ensuremath{E_e}\xspace}

\graphicspath{{figs/}}

\begin{document}

\begin{center}{\Large \textbf{
Opportunities for inclusive diffraction at EIC
}}\end{center}

\begin{center}
	Wojciech S\l{}omi\'nski\textsuperscript{1$\star$},
	N\'estor Armesto\textsuperscript{2},
	Paul R. Newman\textsuperscript{3},
	Anna M. Sta\'sto\textsuperscript{4}
\end{center}

\begin{center}
{\bf 1} Institute of Theoretical Physics, Jagiellonian University, Kraków, Poland
\\
{\bf 2} Instituto Galego de F\'{\i}sica de Altas Enerx\'{\i}as IGFAE, Universidade de Santiago de Compostela, 15782 Santiago de Compostela, Galicia-Spain
\\
{\bf 3} School of Physics and Astronomy, University of Birmingham, UK
\\
{\bf 4} Department of Physics, Penn State University, University Park, PA 16802, USA
\\
* wojtek.slominski@uj.edu.pl
\end{center}

\begin{center}
\today
\end{center}


\definecolor{palegray}{gray}{0.95}
\begin{center}
\colorbox{palegray}{
  \begin{tabular}{rr}
  \begin{minipage}{0.1\textwidth}
    \includegraphics[width=22mm]{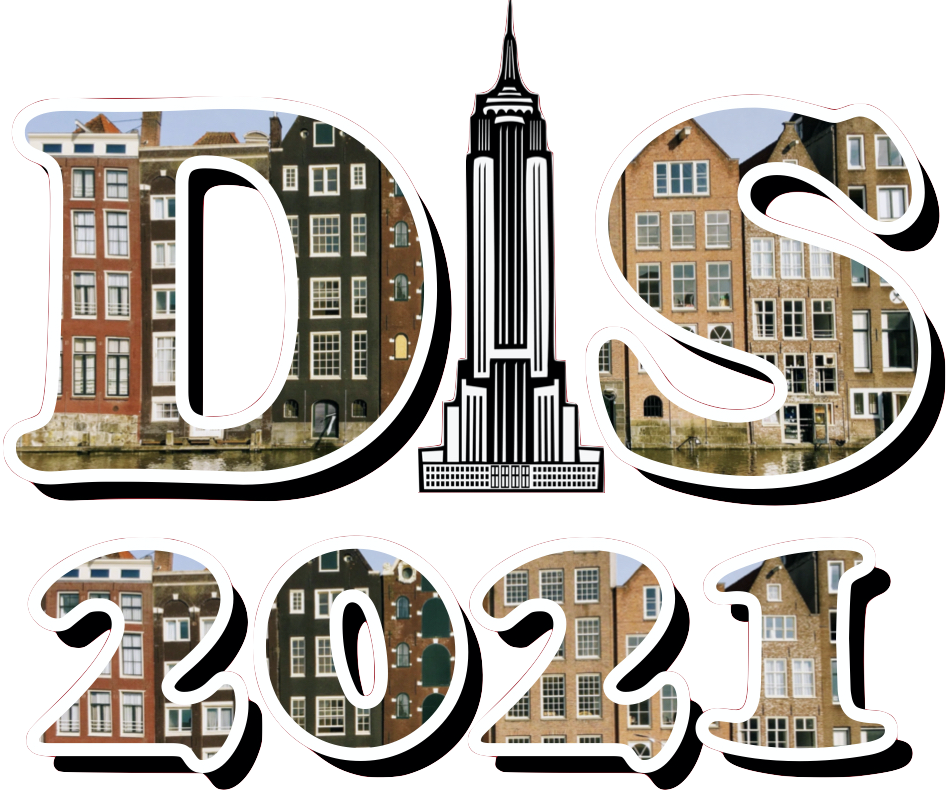}
  \end{minipage}
  &
  \begin{minipage}{0.75\textwidth}
    \begin{center}
    {\it Proceedings for the XXVIII International Workshop\\
		on Deep-Inelastic Scattering and Related Subjects,}\\
    {\it Stony Brook University, New York, USA, 12-16 April 2021} \\
    \doi{10.21468/SciPostPhysProc.?}\\
    \end{center}
  \end{minipage}
\end{tabular}
}
\end{center}


\section*{Abstract}
{\bf
The possibilities for inclusive diffraction in the Electron Ion Collider, EIC, in the US, are analyzed.
We find that thanks to the excellent forward proton tagging, the EIC will be able to access a wider kinematical range of longitudinal momentum fraction and momentum transfer of the leading proton than at HERA. 
This opens up the possibility to measure subleading diffractive exchanges. 
The extended $t$-range would allow the precise extraction of 4-dimensional reduced cross section in diffraction. 
In addition, the varying beam energy setups at the EIC would allow for precise measurements of the longitudinal diffractive structure function.}


\section{Introduction}
\label{sec:intro}

Diffraction in DIS, as observed at HERA, 
consists a large ($\sim 10\%$) fraction of all inclusive events 
in DIS \cite{Adloff:1997sc,Breitweg:1997aa}, see the review \cite{Newman:2013ada} and refs. therein.
In these events the proton 
stays intact or dissociates into a state with the proton quantum numbers.
The experimental signature of such events is a final proton registered in a far forward detector,
and/or the presence of a large rapidity gap (LRG) --- the former case corresponds to a coherent diffraction where the proton does not get diffractively excited.
The hadronic structure of the $t$-channel exchange in these events has been measured at HERA, and parametrized by means of diffractive parton
densities. 

In this presentation we discuss possibilities of measuring inclusive diffraction at EIC~\cite{AbdulKhalek:2021gbh}.
We describe the physical picture behind the performed studies and briefly present the obtained results.
For some more plots the reader is advised to consult the slides of the 
\href{https://indico.bnl.gov/event/9726/contributions/47426/attachments/33770/54339/Slominski_DIS2021.pdf}{talk}.%

\section{Cross section, structure functions and DPDFs}
\label{xsec}

In \Fig{fig:ddis} we show a diagram depicting a neutral current diffractive deep inelastic event
in the one-photon exchange approximation.
Charged currents could also be considered
but in this presentation we limit ourselves to neutral currents.

\begin{minipage}[b]{0.5\columnwidth}
The incoming electron or positron, with four momentum $k$, scatters off the proton, with incoming momentum $p$, and the interaction proceeds through the exchange of a virtual photon with four-momentum $q$.

The distinguishing feature of the diffractive event ${ep\rightarrow eXY}$ is
the presence of 
the diffractive system
of mass $M_X$ and the forward final proton (or its low-mass excitation) $Y$ with four momentum 
\( p' = (E', \vec p_\perp, \xL p_z) \).
\end{minipage}
\hfill
\begin{minipage}[t]{0.4\columnwidth}
\centering
\includegraphics[width=\linewidth,height=45mm,clip,trim=0 0 0 10]{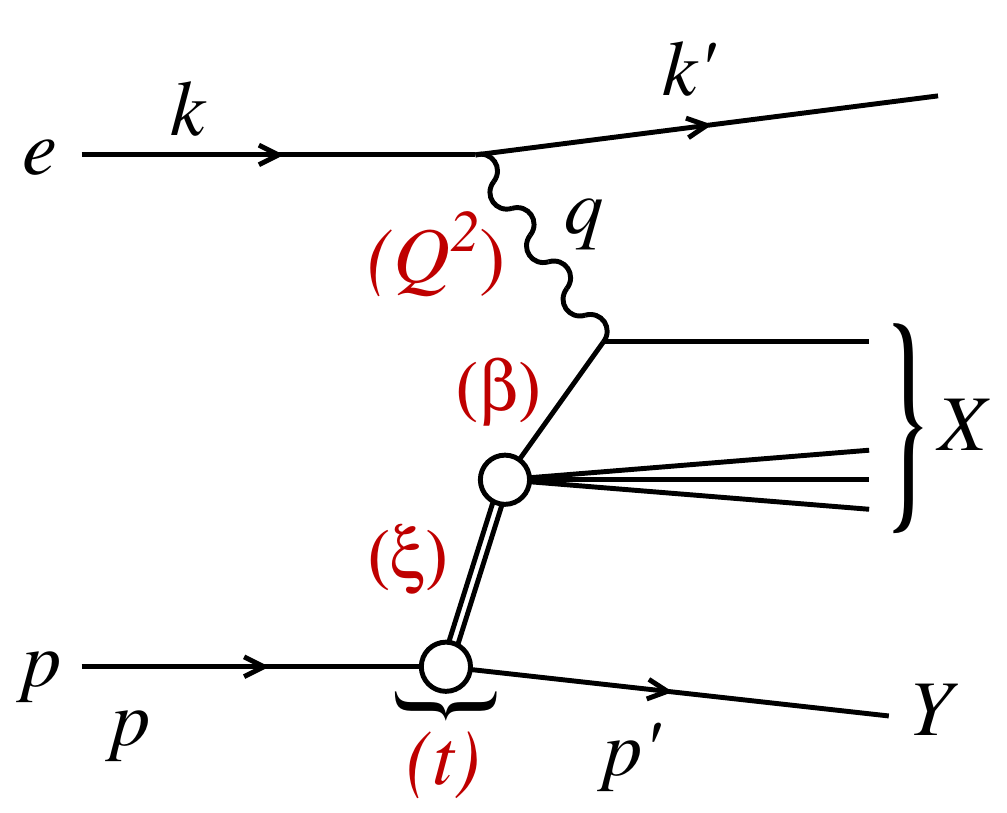}
\captionsetup{width=\linewidth}
\captionof{figure}{A diagram of a diffractive NC event in DIS together with the corresponding variables.}
\label{fig:ddis}
\end{minipage}

 
\bigskip
In addition to the standard DIS variables, diffractive events are also characterized by 
\be
	t=(p-p')^2 = -\frac{p_\perp^2 + (1-\xL)^2 \,m_p^2}{\xL} \,, 
	\qquad \xi=\frac{Q^2+M_X^2-t}{Q^2+W^2}\,, \qquad \beta = \frac{Q^2}{Q^2+M_X^2-t}\, .
\ee
Here  $t$ is the squared
 four-momentum transfer   in the proton vertex, $\xi$ (alternatively denoted by $x_\pom$)  can be interpreted as  the momentum fraction of the `diffractive exchange'   with respect to the hadron,  and  $\beta$ 
is the momentum fraction of the parton with respect to the diffractive exchange. 
 The two momentum fractions combine to give  Bjorken-$x$, $x=\beta \xi$.
 
The physical picture suggested by \Fig{fig:ddis} is that the initial proton splits into a final state $Y$ of momentum $p' \simeq (1-\xi)p$ and the object which is responsible for the diffractive exchange of momentum $\xi p$. The latter, 
in turn, undergoes a DIS-like process to produce the final state $X$. 
The study presented in the following concerns coherent diffraction 
(i.e. the non-dissociative case), where the final state $Y$ is a proton.
Experimentally, this requires tagging of the final proton,
which was performed at HERA using Roman pot insertions to the forward beam-pipe.
At EIC we will have much better forward tagging capabilities
which is discussed in \Sec{xsec4}.
Note that, as compared to HERA, the EIC has rather limited possibilities of the LRG detection,
mainly because of a lower energy range, but also because of the current detector design.

Diffractive cross sections in the neutral current case can be expressed in terms of the  reduced cross 
sections
\be
\frac{d^4 \sigma^{\DD}}{d\xi d\beta dQ^2 dt} = \frac{2\pi \alpha_{\rm em}^2}{\beta Q^4} \, Y_+ \, \sred^{\DD(4)}(\beta,\xi,Q^2,t)\,  ,
\label{eq:sigmared4}
\ee
where
$Y_+= 1+(1-y)^2$. The reduced cross sections are, in turn, expressed by two diffractive structure functions
$F_2^{\DD}$ and $\FL^{\DD}$. 
In the one-photon approximation, the relation reads
\be
\sred^{\DD(4)} = F_2^{\DD(4)}(\beta,\xi,Q^2,t) - \frac{y^2}{Y_+} F_\mathrm{L}^{\DD(4)}(\beta,\xi,Q^2,t) \; .
\ee
The 3-dimensional reduced cross sections,
$\sred^{\DD(3)}(\beta,\xi,Q^2)$, as well as the structure functions, \(F_{2,\mathrm{L}}^{\DD(3)}\),
fulfil analogous relations upon integration over $t$.

Within the standard perturbative QCD approach 
based on colinear factorization 
\cite{Collins:1997sr}
the structure functions read
\begin{equation}
\label{eq:FD4-fac}
F_{2/\mathrm{L}}^{\DD(4)}(\beta,\xi, Q^2,t) =
\sum_i \int_{\beta}^1 \frac{dz}{z}\,
	 C_{2/\mathrm{L},i}\left(\frac{\beta}{z}\right)\, f_i^{\DD(4)}(z,\xi,Q^2,t) \; ,
\end{equation}
where the sum goes over all parton flavors (gluon, quarks), and
the coefficient functions $C_{2/\mathrm{L},i}$ are the same as in  inclusive DIS.
The diffractive parton densities (DPDF) $f_i^{D}$
(that fulfill DGLAP evolution)
are modeled as a sum of two exchange contributions, $\pom$ and $\regg$:
\be
f_i^{\DD(4)}(z,\xi,Q^2,t) =  f^p_{\pom}(\xi,t) \, f_i^{\pom}(z,Q^2)+f^p_{\regg}(\xi,t) \, f_i^{\regg}(z,Q^2) \;.
\label{eq:param_2comp}
\ee
For both of these terms proton vertex factorization is assumed, meaning that the diffractive exchanges can be interpreted as colorless objects called a `Pomeron' or a `Reggeon' with  parton distributions $f_i^{\pom,\regg}(\beta,Q^2)$.
The `Reggeon' term is the only subleading exchange in this model, which was enough to parametrize the HERA data.
The flux factors  $f^p_{\pom,\regg}(\xi,t)$ 
are parametrized using the form motivated by the Regge theory,
\be
 f^p_{\pom,\regg}(\xi,t) = A_{\pom,\regg} \frac{e^{B_{\pom,\regg}t}}{\xi^{2\alpha_{\pom,\regg}(t)-1}} \; .
\label{eq:flux}
\ee

\section{Data simulation}
\label{datsim}

\begin{figure}[h]
\centerline{%
\includegraphics[width=0.32\columnwidth]{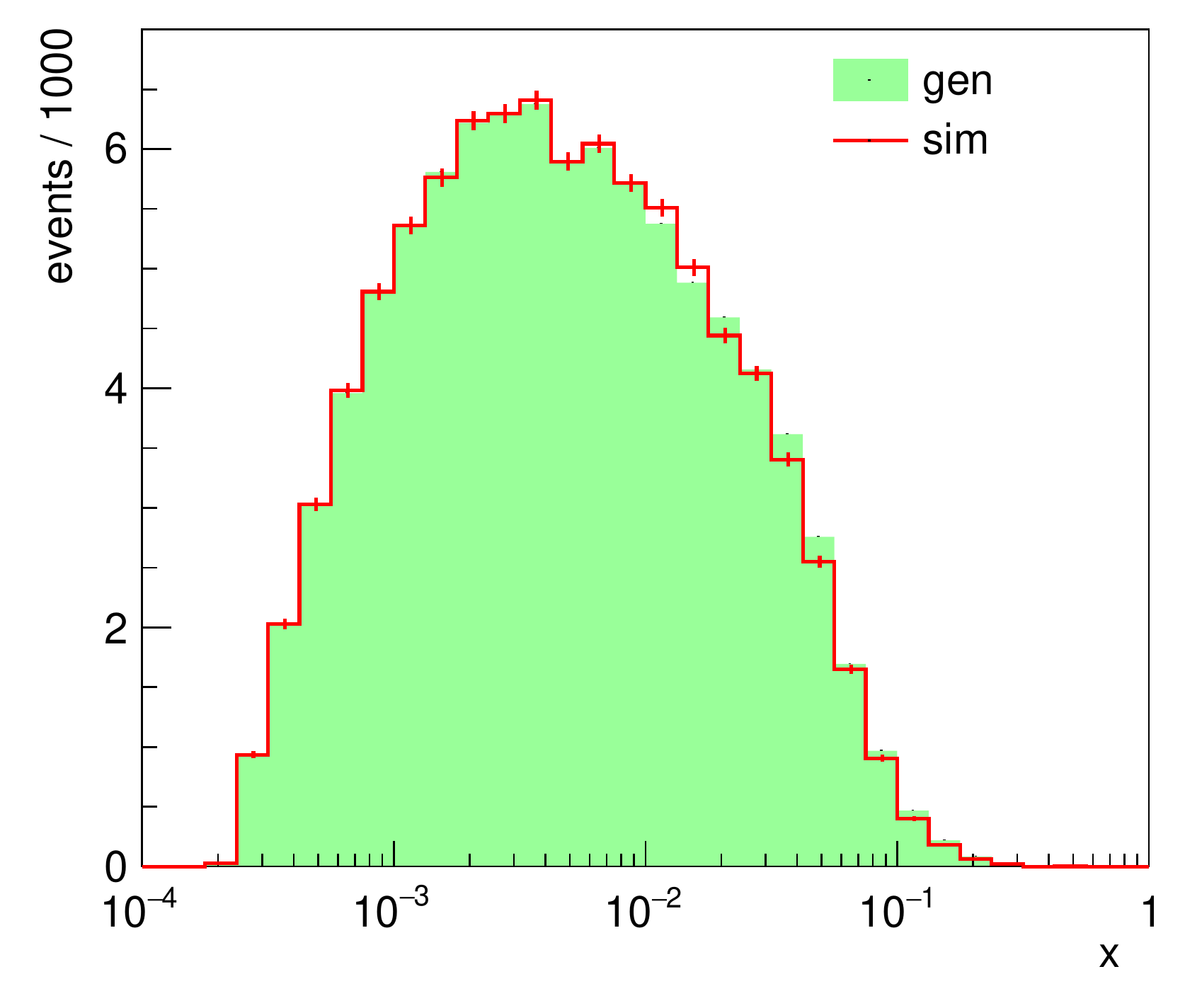}%
\includegraphics[width=0.32\columnwidth]{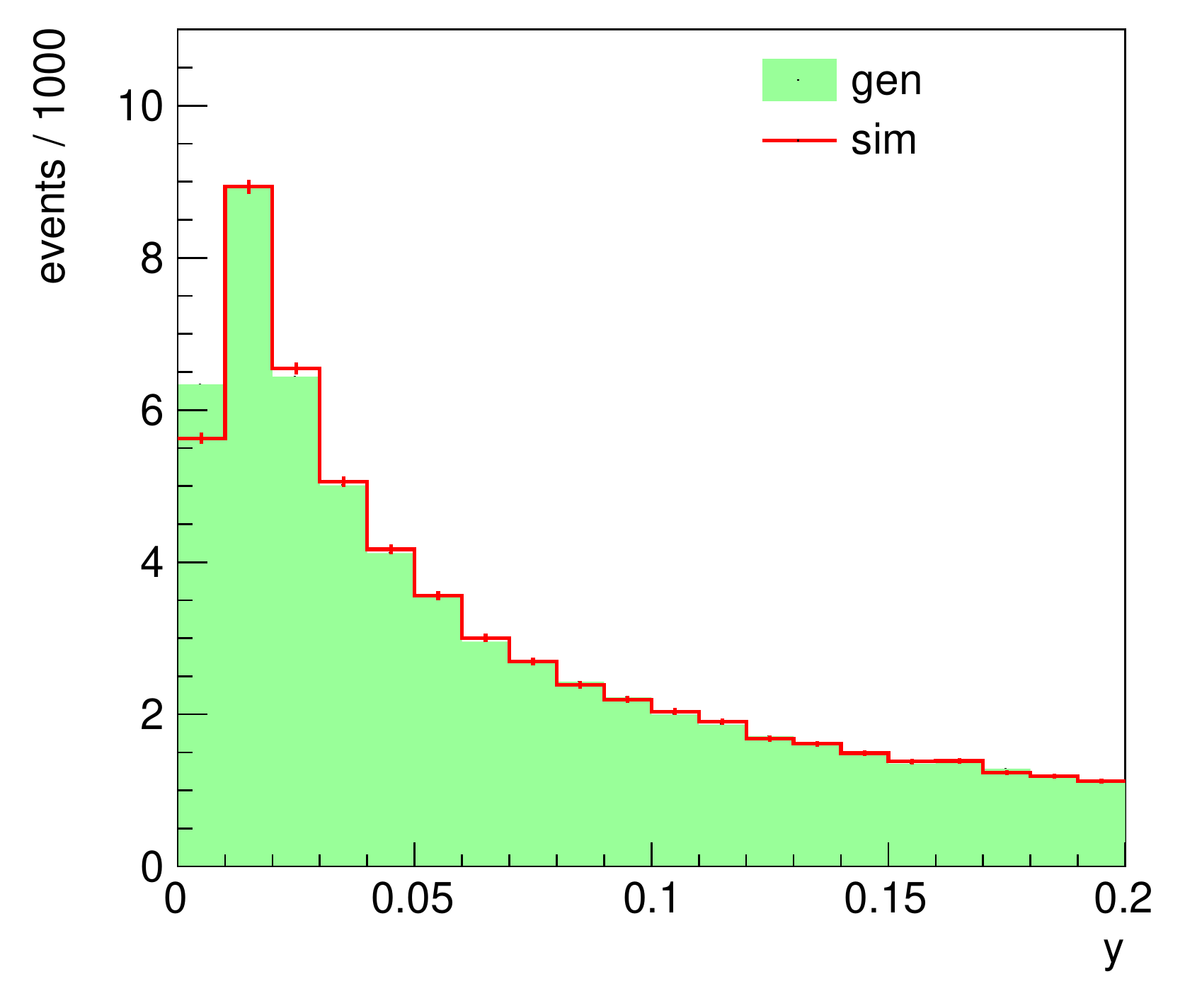}%
\includegraphics[width=0.32\columnwidth]{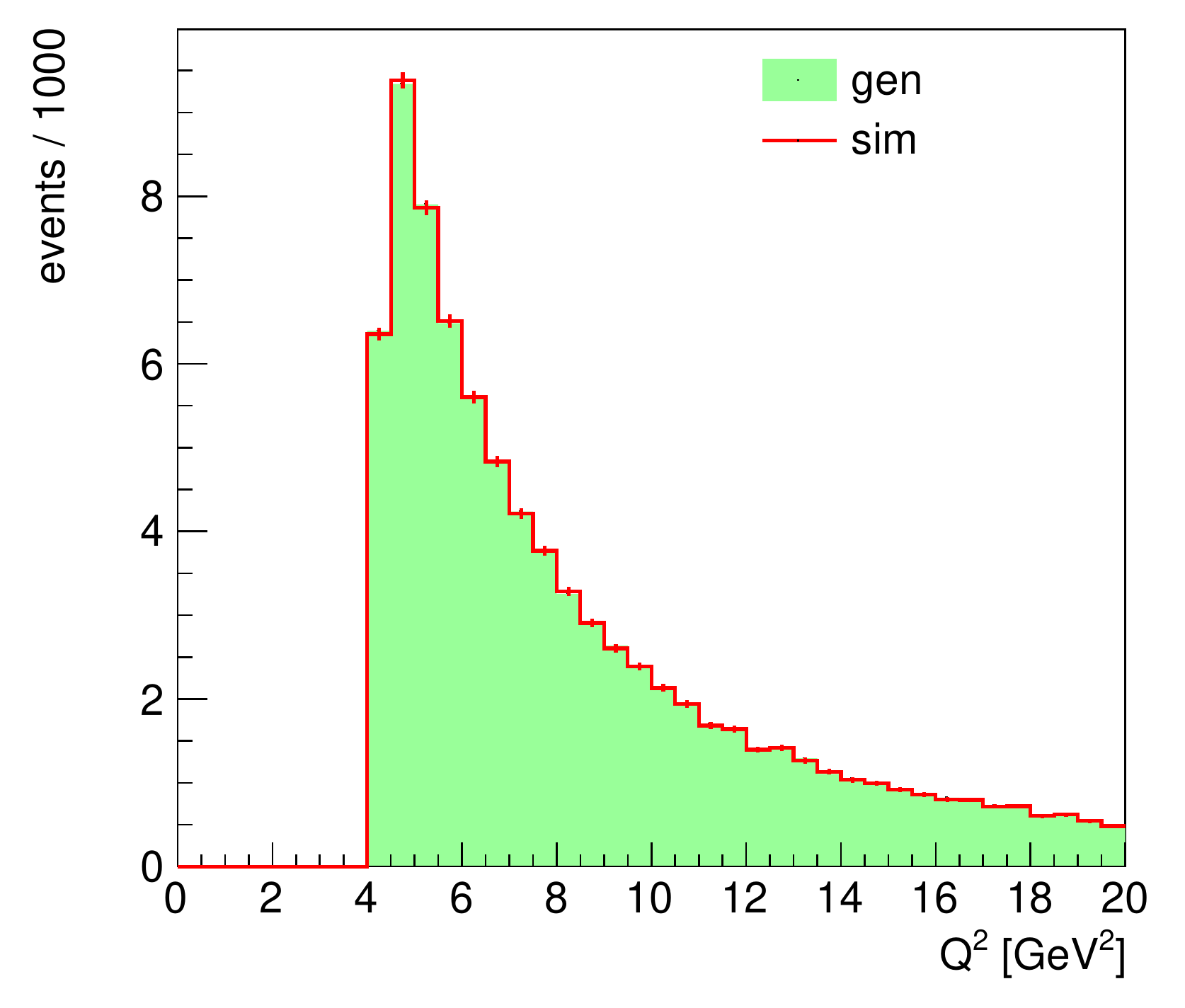}%
}
\caption{\small EIC detector acceptance for the standard DIS variables
for \(E_e\) = 18 GeV and \(E_p\) = 275 GeV.
Green area --- \textsf{Rapgap} generated data;
red histograms --- reconstructed from the detector smeared data.}
\label{fig:xyQ.mc}
\end{figure}
\begin{figure}[!h]
\centerline{%
\includegraphics[width=0.32\columnwidth]{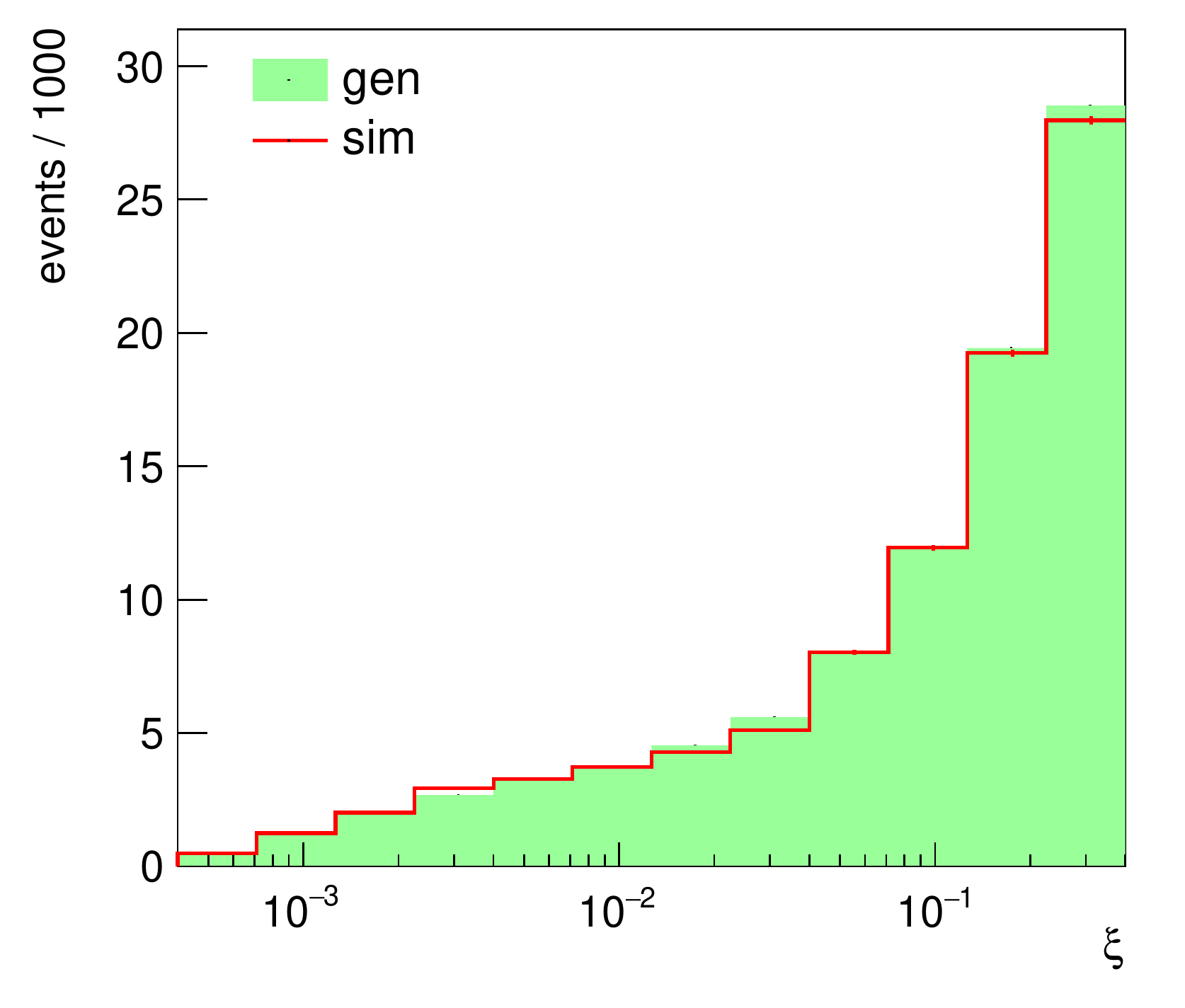}%
\includegraphics[width=0.32\columnwidth]{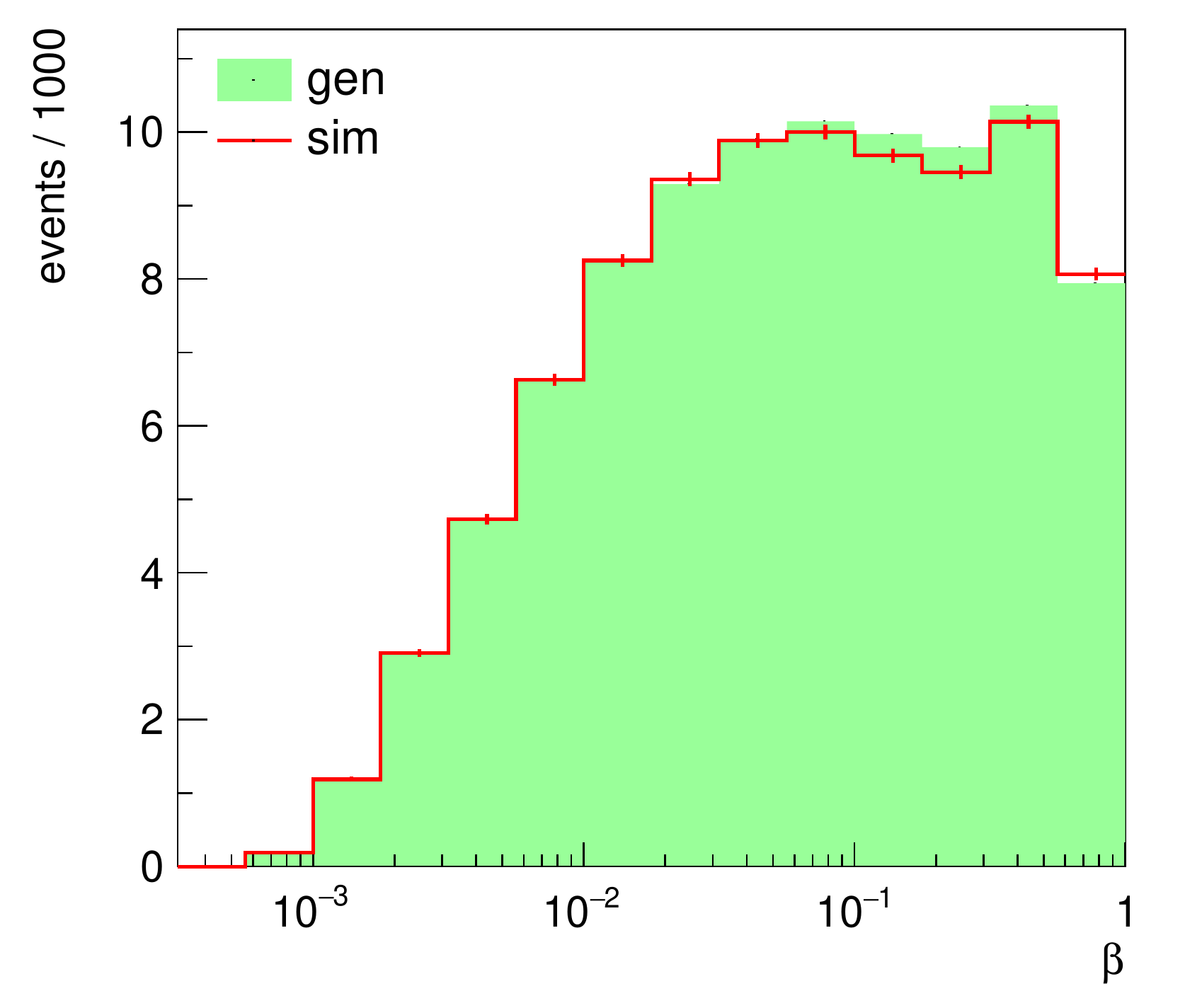}%
}
\caption{\small Same as \Fig{fig:xyQ.mc} but for diffractive variables $\xi$ and $\beta$.}
\label{fig:xib.mc}
\end{figure}

The process of generating pseudo-data
is based on three ingredients: extrapolation, binning selection and systematic error assumption.

The calculation of extrapolated cross section for the $ep$ diffractive DIS at EIC 
follows the model described in \Sec{xsec}. 
The reduced cross sections for selected values of $(\beta,\xi,Q^2,t)$
are obtained by performing
the NLO DGLAP evolution starting from the ZEUS-SJ parametrization \cite{Chekanov:2009aa} of the diffractive PDFs.
For a detailed description see Ref.~\cite{Armesto:2019gxy}.

For the $\sred^{\DD(3)}$ simulations we assume a logarithmic binning with 4 bins per decade in each of \(\beta, Q^2, \xi\).
In order to assure that the EIC detector can provide high quality data for the selected binning
we perform a detailed Monte-Carlo study 
where
the data generated by the \textsf{Rapgap} MC generator (\cite{Jung:1993gf}, see also {\tt https://rapgap.hepforge.org/}) are passed through the detector simulation (using the EICsmear code~\cite{AbdulKhalek:2021gbh}).
Several kinematic reconstruction methods are considered leading finally to an optimized method consisting in taking an average from
several of them weighted by their resolution.
Within this optimized method
a very good acceptance is achieved, as shown in Figs.~\ref{fig:xyQ.mc} and \ref{fig:xib.mc}.

Once the bins are fixed we calculate the cross sections 
and add random Gaussian smearing according to the estimated \estat and \esys added in quadrature.
\estat is calculated from the cross section value and the assumed integrated luminosity of 2 or 10 fb\(^{-1}\);
\esys is taken in the  1--5\% range. Except for very large $Q^2$ (above $500\,\GeV^2$), 
the errors of simulated data are dominated by \esys.



\section{Diffractive PDFs from fits to pseudo-data}

\begin{figure}[!h]
\centering
\begin{subfigure}[b]{0.5\linewidth}
	\includegraphics[width=\linewidth,clip,trim=0 260 0 43]{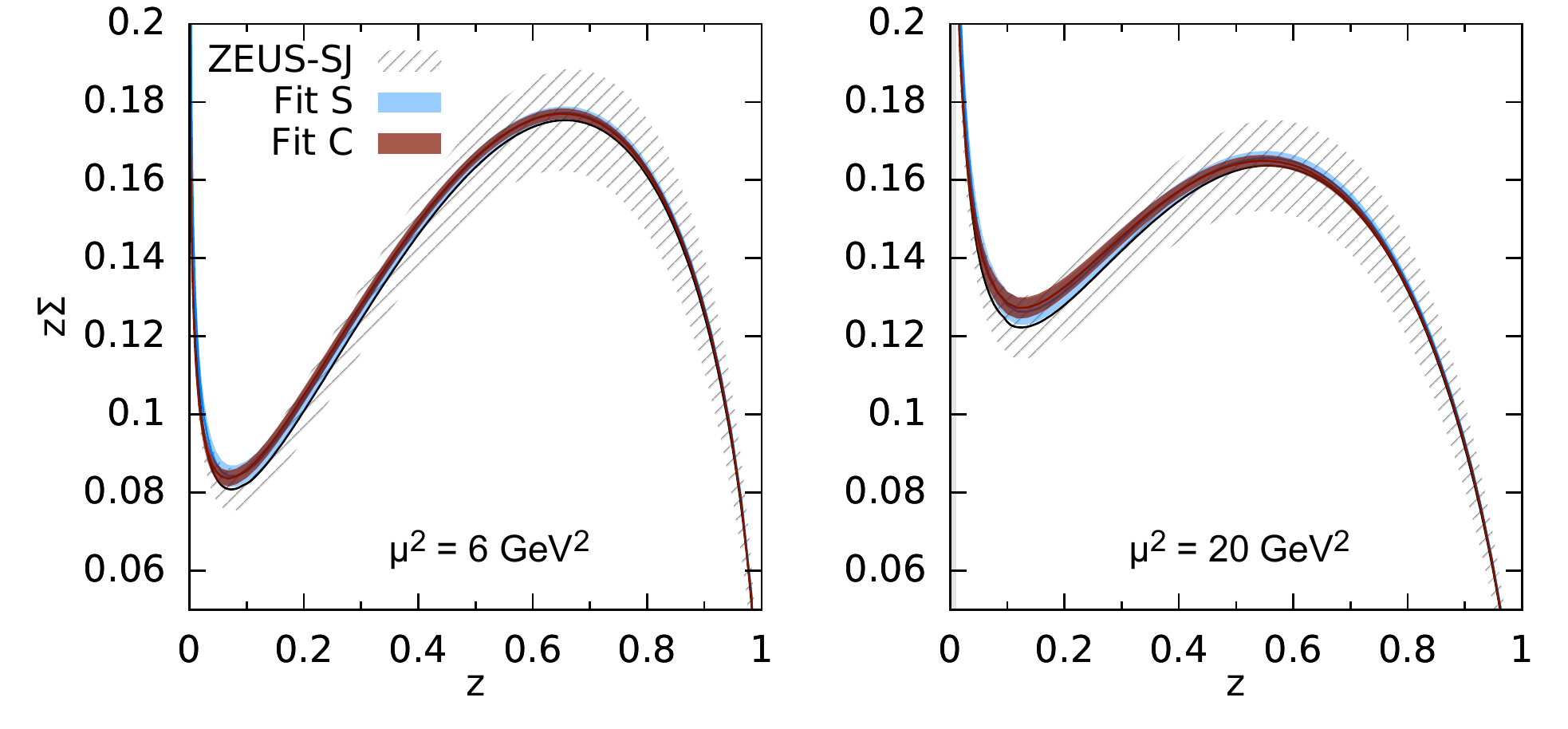}%
	 \caption{\small quarks}
\end{subfigure}%
\begin{subfigure}[b]{0.5\linewidth}
	\includegraphics[width=\linewidth,clip,trim=0 260 0 43]{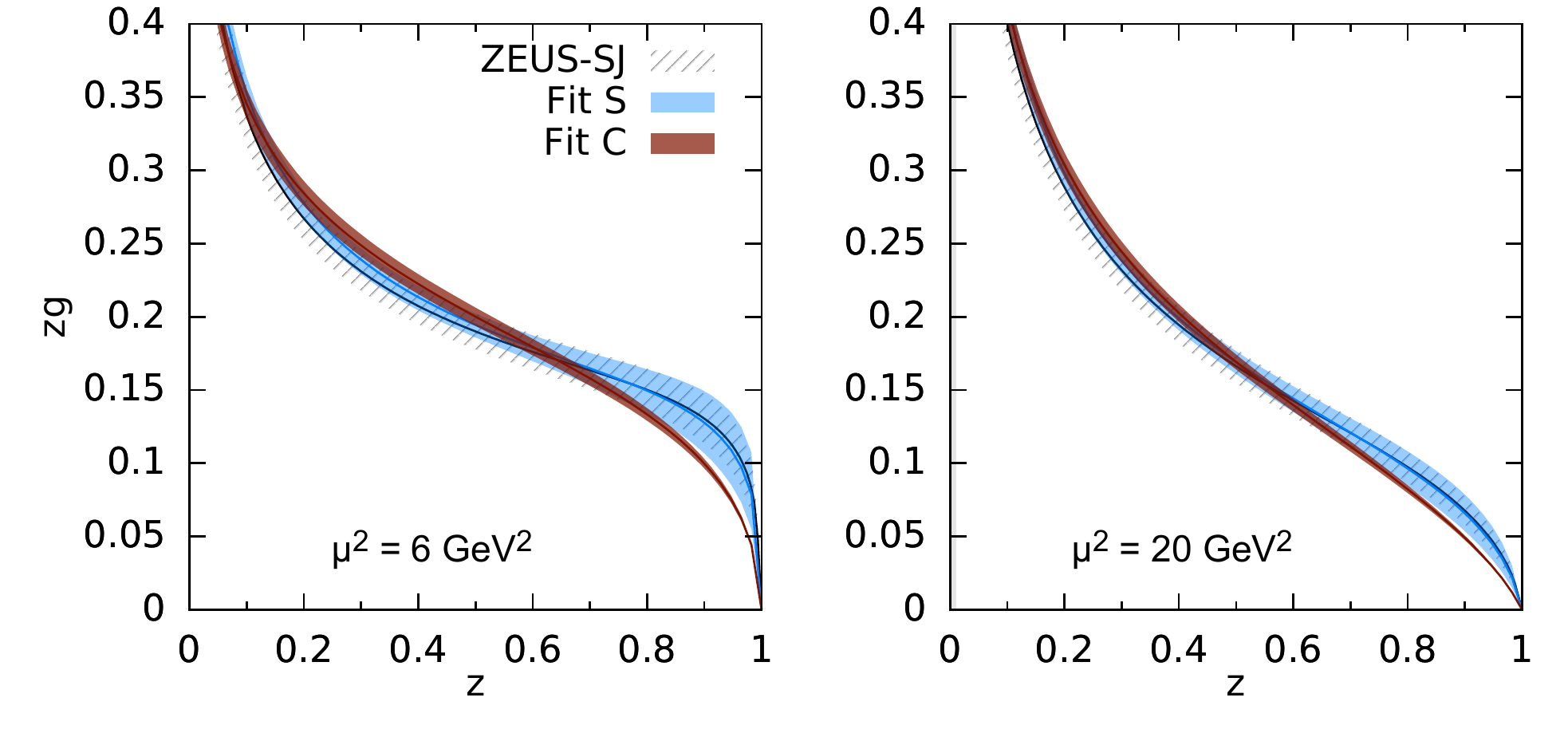}%
	 \caption{\small gluon}
\end{subfigure}%
\caption{\small DPDFs from fits to the pseudo-data for \(E_e\) = 18 GeV and \(E_p\) = 275 GeV.}
\label{fig:xi-sr3}
\end{figure}

The diffractive parton distribution functions were extracted from
the fits to pseudo-data following the procedure explained in Ref.~\cite{Armesto:2019gxy}. 
As compared to HERA we observe much smaller uncertainty for
the quark DPDFs, at least at high values of the longitudinal momentum fraction $z$ ($z \ge \beta$).
The extraction of gluon DPDFs
from the inclusive data only
 requires access to very low $z$ and thus
we get no improvement with respect to HERA.


\section{Longitudinal structure function}

\begin{figure}
\centerline{%
	\includegraphics[width=0.75\columnwidth,clip,trim=0 12 0 192]{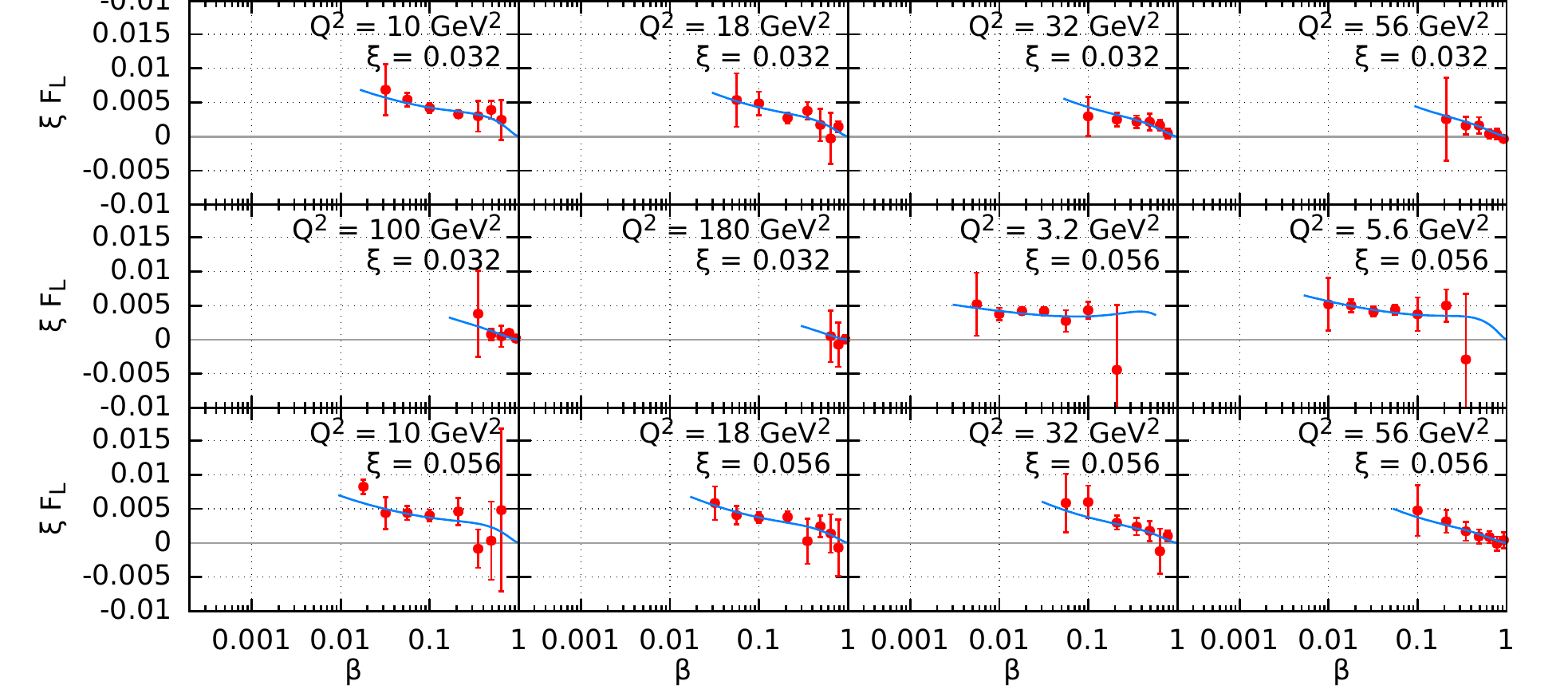}%
}
\caption{\small $\FL^{\DD(3)}$ from fits to pseudo-data simulated at 17 center-of-mass energies for
\esys = 2\% and integrated luminosity of 10 fb\(^{-1}\) --- a small subsample.}
\label{fig:FL}
\end{figure}

At fixed $(\beta,\xi,Q^2)$ the reduced cross section $\sred^{\DD(3)}$ depends on the center-of-mass energy
squared, $s$, via the inelasticity $y = Q^2/(s\beta\xi)$,
\be
\label{eq:sr-F2L}
\sred^{\DD(3)} = F_2^{\DD(3)}(\beta,\xi,Q^2) - Y_\mathrm{L}\, F_\mathrm{L}^{\DD(3)}(\beta,\xi,Q^2) \,,
\ee
where $Y_\mathrm{L} = y^2/Y_+$.

This, rather weak, dependence comes from the non-zero longitudinal structure function $\FL^{\DD(3)}$ and is practically measurable at high $y$ only.
Precise measurements at several center-of-mass energies can provide data on $F_{2,\mathrm{L}}^{\DD(3)}$
by fitting \Eq{eq:sr-F2L} as a function of $Y_\mathrm{L}$.

To explore the possibility of such a measurement we considered three electron energies
\Ee = {5, 10, 18} GeV
and 6 proton energies \Ep = {41, 100, 120, 165, 180, 275} GeV,
resulting in 17 different center-of-mass energies.
For the assumed binning we obtained 469 $(\beta,\xi,Q^2)$ bins, each containing at least four $\sred^{\DD(3)}$ data points.
After fitting we got $\FL^{\DD(3)}$ vs. $\beta$ plots in 76 $(\xi,Q^2)$ bins.
Some examples are shown in \Fig{fig:FL}.

\section{4-dimensional diffractive cross section}
\label{xsec4}

\begin{figure}[hb]
\centerline{%
\includegraphics[width=0.85\columnwidth,clip,trim=0 0 0 6]{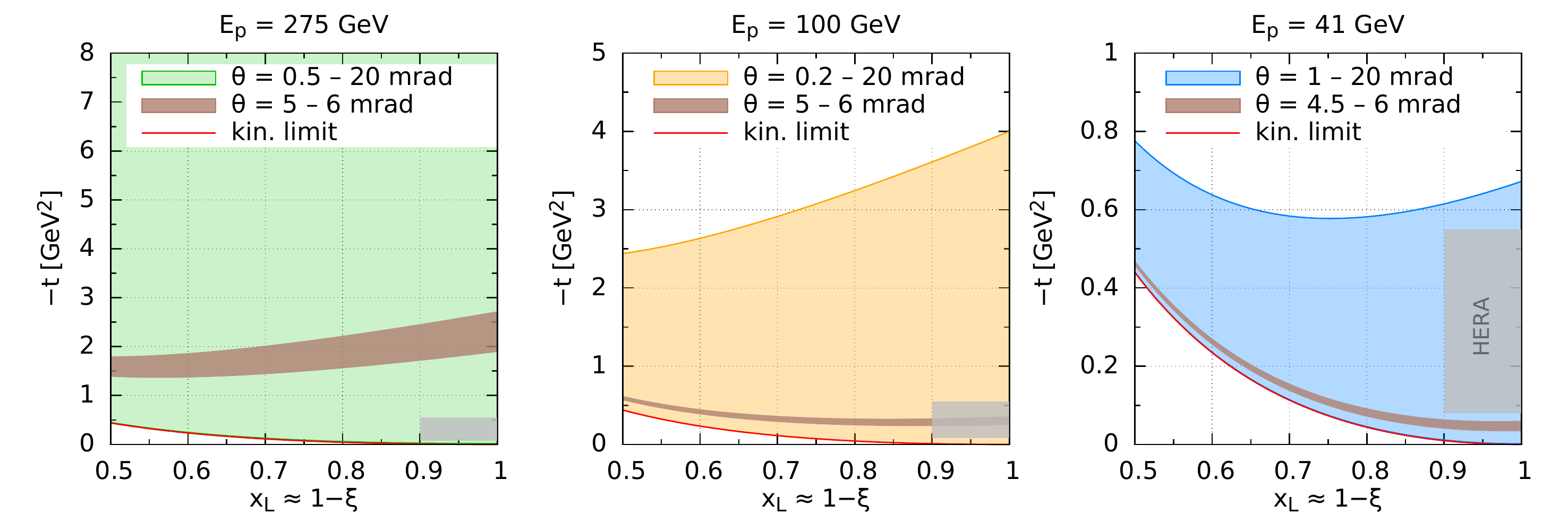}%
}
\caption{\small Final proton tagging.
 \(\xL, t\) range of the proton tagged by the EIC detector for three proton energies, 275 GeV, 100 GeV and 41 GeV.
The brown strip marks a small ($\sim$ 1 mrad) region not covered by the current detector design.}
\label{fig:xL-t}
\end{figure}

The EIC detector provides an excellent tagging of the final proton in a wide range of $t$ and \xL.
In \Fig{fig:xL-t} we show the kinematic range covered at three proton energies.
In each plot the range covered by HERA is marked with a gray rectangle,
 and we see that EIC opens a new area for measurements of the leading proton. 

\begin{figure}[!hb]
\centerline{%
	\includegraphics[width=0.45\columnwidth,clip,trim=0 0 0 22]{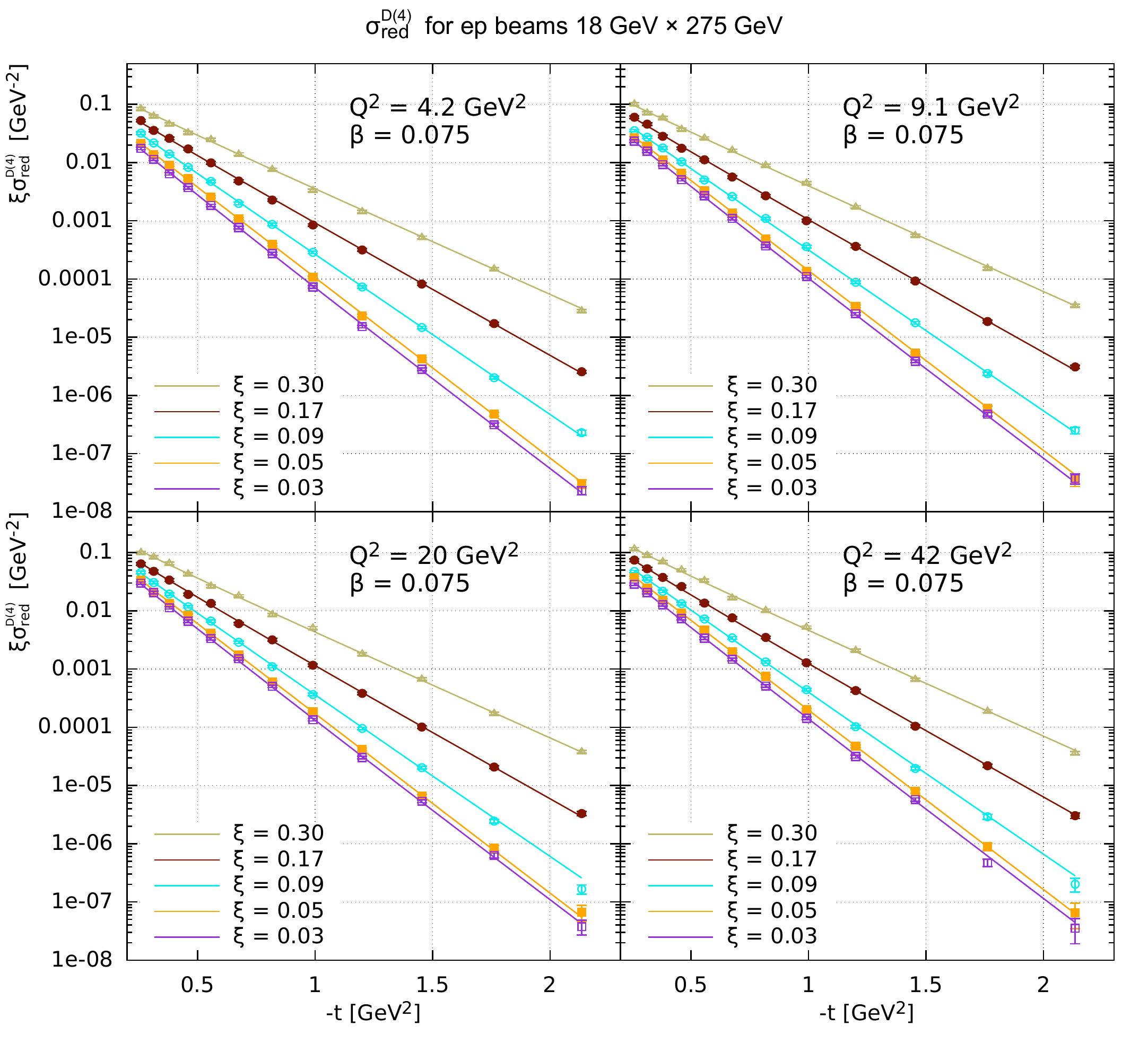}%
	\includegraphics[width=0.45\columnwidth,clip,trim=0 0 0 22]{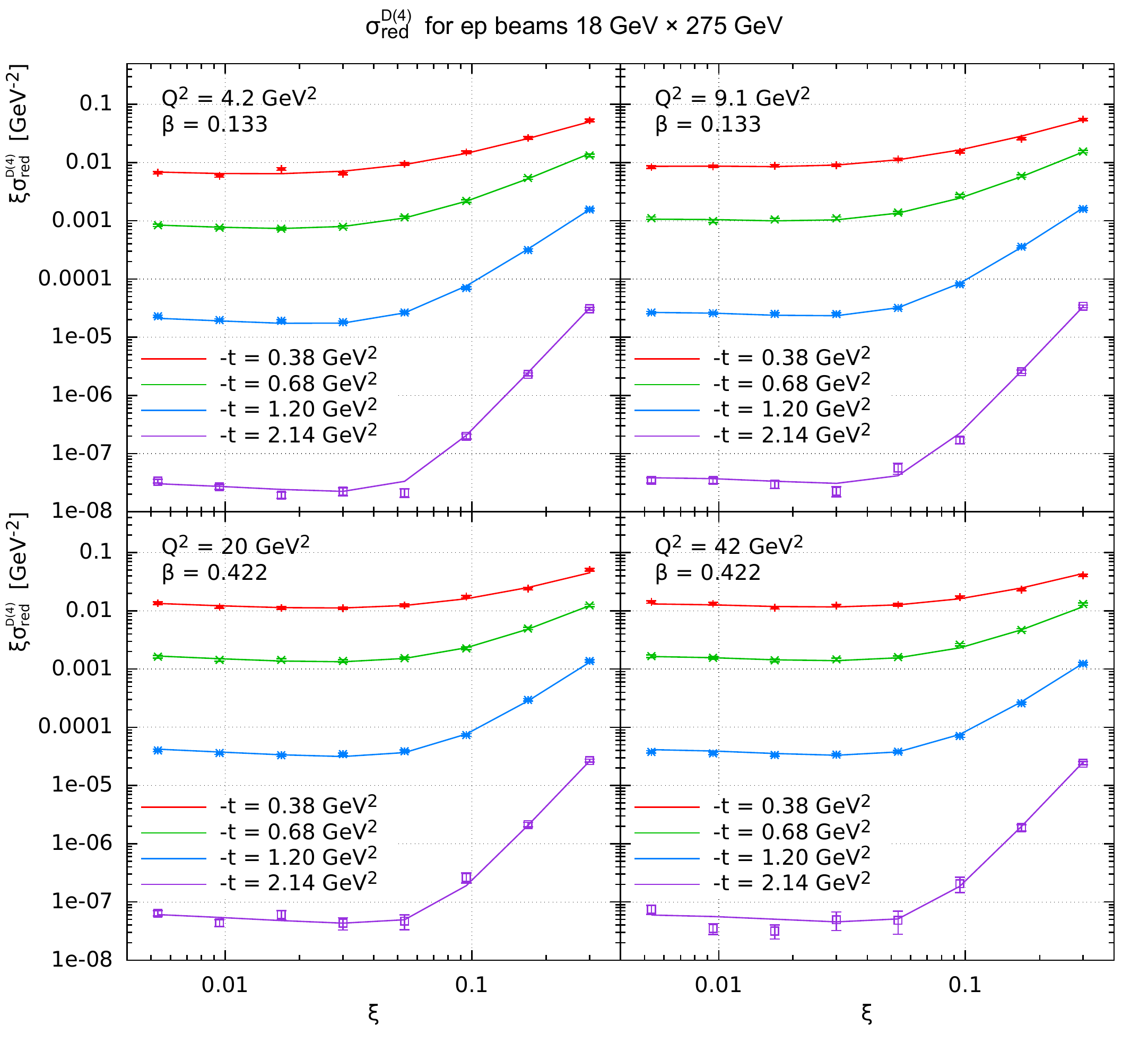}%
}
\caption{\small \(\sred^{\DD(4)}\) dependence on $t$ and $\xi$ in selected \((Q^2, \beta)\) bins.}
\label{fig:xi-t-sr4}
\end{figure}

In order to simulate data for the reduced cross section $\sred^{\DD(4)}$
we have first performed the study of acceptance for the assumed binning of the variables related to the proton tagging,
\(t,\xL,\pT\), as described in \Sec{datsim}.
We have confirmed a very good agreement of the reconstructed vs. simulated data.

The results of simulations for \esys = 5\% and integrated luminosity of 10 fb\(^{-1}\) are shown in \Fig{fig:xi-t-sr4}.
It can be seen that the dependence of $\sred^{\DD(4)}$
on both $t$ and $\xi$ can be measured with a very good precision.
In particular, the double slope structure of the $\xi$ dependence 
allows for a clear separation of the leading and subleading exchange contributions
according to \Eq{eq:param_2comp}.

\section{Conclusion}

We have studied the capabilities of the EIC  to measure diffraction in the inclusive DIS.
We observe an excellent final proton tagging and a very good acceptance and resolution for the diffractive DIS variables.

Using simulated data for $\sred^{\DD(3)}$ and $\sred^{\DD(4)}$ we point out the possibilities of a precise extraction of the quark DPDFs,
a measurement of $\FL^{\DD(3)}$ in a wide kinematic range and a precise determination of the subleading exchange contribution from the $t$-dependence studies.

\section*{Acknowledgements}
We thank Hannes Jung
for useful discussions.
NA acknowledges financial support by Xun\-ta de Galicia (Centro singular de investigaci\'on de Galicia accreditation 2019-2022); the "Mar\'{\i}a de Maeztu" Units of Excellence program MDM2016-0692 and the Spanish Research State Agency under project FPA2017-83814-P;
European Union ERDF; the European Research Council under project
ERC-2018-ADG-835105 YoctoLHC; MSCA RISE 823947 "Heavy ion collisions: collectivity and precision in saturation physics"
(HIEIC); and European Union's Horizon 2020 research and innovation programme under
grant agreement No. 824093.
AMS is supported by the U.S. Department of Energy grant No. DE-SC-0002145 and in part by National Science Centre in Poland,
grant 2019/33/B/ST2/02588.

\bibliography{mybib.bib}

\nolinenumbers

\end{document}